\documentclass[prd, twocolumn, floatfix, superscriptaddress]{revtex4}

\usepackage{epsfig} 
\usepackage{amsmath}
\usepackage{amsbsy,color}

\newcommand{\beq}{\begin{equation}}
\newcommand{\eeq}{\end{equation}}
\newcommand{\beqa}{\begin{eqnarray}}
\newcommand{\eeqa}{\end{eqnarray}}
\newcommand{\om}{\Omega_m}

\newcommand{\nerr}{N_{\rm err}}
\newcommand{\nsn}{N_{\rm SN}}
\newcommand{\sint}{\sigma_{\rm int}}
\newcommand{\zmax}{z_{\rm max}}
\newcommand{\lmax}{\lambda_{\rm max}}

\newcommand{\aap}{{Astron.~Astrophys.}}
\newcommand{\apjl}{{Astrophys.~J.~Lett.}}

\newcommand{\mnras}{{Mon.~Not.~R.~Astron.~Soc.}}
\newcommand{\jcap}{{J.~Cosmol.~Astropart.~Phys.}}

\begin{document} 

\title{Correlated Supernova Systematics and Ground Based Surveys} 
\author{Alex G.\ Kim}
\affiliation{Lawrence Berkeley National Laboratory, Berkeley, CA 94720 USA}
\author{Eric V.\ Linder} 
\affiliation{Lawrence Berkeley National Laboratory, Berkeley, CA 94720 USA}
\affiliation{University of California, Berkeley, CA 94720, USA}
\affiliation{Institute for the Early Universe, Ewha Womans University, Seoul, Korea} 
\date{\today}

\begin{abstract} 
Supernova distances provide a direct probe of cosmic acceleration, 
constraining dark energy.  
This leverage increases with survey redshift depth at a rate bounded 
by the systematic uncertainties.  We investigate the impact of a 
wavelength-dependent, global correlation model of systematics in 
comparison to the standard local-redshift correlation model.  This can 
arise from subclass uncertainties as features in the supernova spectrum 
redshift out of the observer 
photometric filters or spectral range.  We explore the impact of such a 
systematic on ground-based supernova surveys such as Dark Energy Survey 
and LSST, finding distinctive implications.  Extending the wavelength 
sensitivity to 1.05$\,\mu$m through 
``extreme red'' CCDs can improve the dark energy figure of merit by 
up to a factor 2. 
\end{abstract} 

\maketitle

\section{Introduction \label{sec:intro}}

Type Ia supernovae (SNe) are standardizable distance indicators that can 
be seen to high redshift, over at least 70\% of the lookback time of the 
universe. In a given galaxy, they occur on $10^2$--$10^3$ year time scales, 
so for $10^{10}$ galaxies in the visible universe this implies one SN 
per second.  A three-year deep survey could theoretically detect $10^8$ 
SNe so if the distance to each were known to 20\% (0.46 mag; i.e.\ no 
standardization applied, just identification as Type Ia), then one 
could attain 0.01\% distance measurements in 25 equal-error distance 
bins statistically.  In this sense SNe are the most powerful cosmological 
probe statistically. 

However, as for any probe, the key issue comes down to systematic 
uncertainties not statistics.  For SNe, such uncertainties include 
flux calibration, dust corrections, and population drift.  Numerous 
articles have addressed one or another of these, for various 
experimental designs \cite{systematicspapers}.  Some uncertainties 
can be removed by using spectrophotometry rather than filter measurements; 
others can be ameliorated by observations spanning a long wavelength range; 
others by ``like to like'' comparison of time-series or spectral features 
\cite{coping,linlike}.  

To allow general comparison of different SN surveys, a phenomenological 
systematic model offers a compact way of summarizing the specifics of 
the measurement and analysis chain that delivers distances.  
It also provides a basis with which to identify previously 
unknown sources of uncertainty.  Specifically, the model can contain 
parameters describing the distribution of supernova subsets, e.g.\ 
magnitude offsets, that when fit can be checked for deviations or 
inconsistencies. 

A key feature 
of such a model would be a systematics floor, reflecting that the large 
number of SNe mentioned above should not in reality give an asymptotically 
small precision.  This is fundamentally different from ``self calibration'' 
-- adopting a fixed parametric form and simultaneously fitting the error 
parameters along with the cosmology\footnote{We 
note that this is effectively what happens with SN light curve fitting 
parameters.  Some approaches fit for the SN parameters first, 
deriving distances, and then use those distances to fit for cosmology, 
but it can be more powerful to carry out the cosmology fit in one step 
\cite{kimmiq,facc}.}.  
Self calibration only holds if there 
is no uncertainty in the form, only the parameters.  This is a weaker 
meaning of systematic uncertainty, and in general we do not have strong 
confidence in the form for population drift, for example.  Therefore we 
consider a systematics model that explicitly includes a floor.

A well-used systematics model was introduced by \cite{linhut}, consisting 
of local correlations between SNe within redshift bins of width 0.1.  Large 
numbers of SNe in a bin reduce the distance error to a finite floor, 
but continued cosmological leverage could be obtained by adding independent 
SNe, those in another redshift bin\footnote{One 
can view the correlation of SNe within the redshift bin as a block 
diagonal component consisting of each of the individual SNe within the 
bin.  That is, one does not have to bin the SNe per se, just correlate 
them.  Having a $1700\times 1700$ error matrix of individual SNe, with each 
group of 100 between $z=0.1(i-1)$ and $0.1i$ where $i=1,2,\dots 17$ 
say, correlated, hence giving a block diagonal structure, gives  
mathematically equivalent results to a $17\times 17$ diagonal matrix 
with the statistical error reduced by the square root of the number of SNe 
within the group.}.  
The floor is given in magnitudes by 
\begin{equation}
\sigma\equiv 5\log\frac{\delta d}{d}=\sigma_{\rm sys}\,\frac{1+z}{1+\zmax}\,, 
\label{LH:eqn}
\end{equation} 
where $d$ is the luminosity distance, $z$ is the redshift of the bin 
center, and $\zmax$ is the maximum redshift of the survey; 
we refer to this as the LH systematics model.  An important feature is 
that the floor does not vanish even for $z=0$.  
The cosmological 
parameter estimation results are fairly insensitive to the bin width 
(scaling the error amplitude as the inverse square root of the width) 
within the range $\Delta z=0.05-0.4$ \cite{vernier}. 

However, we want to explore beyond this local correlation model, where 
only SNe within a restricted redshift range are correlated with each 
other.  One can imagine that some systematics might correlate all SNe, 
or all SNe above a certain redshift (e.g.\ if they rely on near-infrared 
observations).  This can give a more global correlation, and we consider 
a model for that here. 

In Sec.~\ref{sec:corr} we present a motivated model for correlated 
systematic uncertainties, and in Sec.~\ref{sec:cos} propagate its effects 
through to dark energy cosmology fitting, especially the dark energy 
equation of state figure of merit.  We investigate in particular 
the influence of various survey and supernovae characteristics 
on the figure of merit.  Section~\ref{sec:other} extends this to 
error fitting and to error models describing population drift.  The 
results are summarized in Sec.~\ref{sec:concl}.

\section{Correlated Systematic \label{sec:corr}}

An error model with every supernova potentially correlated with every 
other supernova is general, but one should use some physical motivation 
to guide the form of the correlations\footnote{Moreover, for an arbitrary 
set of correlations the covariance matrix is not guaranteed to give a 
positive definite result.}.  
Our systematic model is based on SN spectral characteristics and 
says that errors, once they 
enter the measurements, affect all SNe above the entry redshift.  
We consider several different error contributions, each with their 
own entry redshift.  This correlates that portion of the systematic 
uncertainty completely among all higher redshift SNe.  The error matrix 
can be viewed as being ``laid down'' in steps, with the first error 
source contributing $\sigma_1^2$ to all elements of the error matrix, 
the second error source adding $\sigma_2^2$ to the matrix block 
corresponding to higher redshift SNe, and so on for even higher redshift 
SNe.  Effectively, the error matrix has a ``wedding cake'' form, 
climbing in steps from the upper left to lower right.  

Such an error structure naturally arises as supernova light redshifts 
out of the range of observational sensitivity.  For example, the SN Ia 
subclassification scheme of Branch et al.\ (2006) \cite{branch06} 
identifies four groups based on the equivalent widths of absorption features 
at rest-frame 5750 and 6100 \AA\ (from Si II), and the presence of 
high-velocity Ca II IR3 
absorption at 8000 \AA.  All three lines are available for $z<0.12$ 
candidates from ground-based spectroscopic observations within the 
$0.35\, \mu\mbox{m} < \lambda < 0.9\, \mu\mbox{m}$ atmospheric window.  

The spectral features contribute to the standardization of the SNe 
as distance indicators; this can be viewed as calibrating the SN 
absolute magnitude $M$.  
As one by one the spectral features become inaccessible beyond $z=0.12$, 
0.47, and 0.57, supernovae at higher redshifts incur additional error 
floors from imperfect marginalization over each lost calibration 
parameter.  This 
produces an error model where progressively higher redshift bins share 
additive correlated uncertainties.  Note that even with measurements of 
these spectral features, the ensemble of supernova distances is still 
expected to have a correlated error floor due to systematic limitations 
of such a classification scheme in general.  

The principles of a wavelength dependent, globally correlated systematic 
can be extended to the larger set of spectral features cited in the 
literature as indicators of SN Ia diversity \cite{spectraldiversity}. 
Moreover, since relevant features generally have large equivalent 
widths (due to the high velocities in the exploding supernova), they 
may show up in broadband photometry as well.  
The bounded observer window also induces
redshift-dependent correlations in broadband
photometric analysis from propagated uncertainties of
SN and dust models at common restframe wavelengths. 
We therefore investigate 
application of such a correlated systematics model to surveys 
involving either broadband photometry or spectroscopic followup. 

Mathematically, the SN absolute magnitude is taken to be determined as 
a function of measurements of $N$ spectral indicators.  At higher 
redshifts, some of these, say $N-i$ redder 
indicators, are unavailable.  This leads to increased magnitude uncertainty, 
and a covariance in magnitude (and hence distance) errors of 
\begin{equation}
\left<\delta M_{a}\,\delta M_{b}\right> = \sigma^2_{{\rm int},n_a} 
\delta_{ab} + V_{ab} \,, 
\label{MCov:eqn}
\end{equation}
where the intrinsic dispersion enters as
a diagonal 
contribution $\sint^2$ (generally dependent on the procedure used to
determine $M$ for each supernova), and
\begin{equation}
V_{ab}= \sum_{i=\max{\left(n_a,n_b\right)}}^N\sigma^2_{N-i+1}\,.
\label{V:eqn}
\end{equation}
The indices $a$, $b$ label individual SNe and $n_a$ is the number 
of spectral features observed for supernova $a$.  The more spectral 
information, the fewer terms contribute to the sum and the smaller the 
overall uncertainty.  However note that we always keep a base variance 
$\sigma_1^2$. 

An example covariance matrix for three error sources (and five SNe in 
increasing redshift order) is shown below: 
\beq 
V= 
\begin{pmatrix} 
\sigma_1^2 & \sigma_1^2 & \sigma_1^2 & \sigma_1^2 & \sigma_1^2\\ 
\sigma_1^2 & \sigma_1^2 & \sigma_1^2 & \sigma_1^2 & \sigma_1^2\\ 
\sigma_1^2 & \sigma_1^2 & \sigma_1^2+\sigma_2^2 & \sigma_1^2+\sigma_2^2 & 
\sigma_1^2+\sigma_2^2\\ 
\sigma_1^2 & \sigma_1^2 & \sigma_1^2+\sigma_2^2 & \sigma_1^2+\sigma_2^2 & 
\sigma_1^2+\sigma_2^2\\ 
\sigma_1^2 & \sigma_1^2 & \sigma_1^2+\sigma_2^2 & \sigma_1^2+\sigma_2^2 & 
\sigma_1^2+\sigma_2^2+\sigma_3^2 
\end{pmatrix} \,. \label{eq:vsample} 
\eeq 

Another way to achieve the same result is to view the table of error 
contributions for each SN as a $N_{\rm SN}\times \nerr$ matrix.  The 
product of this with its transpose will automatically give a positive 
definite, symmetric matrix.  That is, let 
\beq 
S= 
\begin{pmatrix} 
\sigma_1 & 0 & 0\\ 
\sigma_1 & 0 & 0\\ 
\sigma_1 & \sigma_2 & 0\\ 
\sigma_1 & \sigma_2 & 0\\
\sigma_1 & \sigma_2 & \sigma_3 
\end{pmatrix} \,.
\eeq 
Then the product $V=SS^T$ gives Eq.~(\ref{eq:vsample}).  Here we see 
that each row gives the systematic error summary of a SN, and each 
column represents a different systematic error contribution.  So all 
SNe suffer from uncertainty type 1, with magnitude $\sigma_1$ (representing 
that even with the identified spectral features there will be residual, 
unidentified systematics); only the 
last three SNe (which could represent higher redshifts) suffer from 
type 2 uncertainty, and only the last SN (e.g.\ highest redshift) has 
type 3 uncertainty.  Thus, the fourth SN gets contributions from type 
1 and 2 uncertainties, while the fifth SN has systematics of types 1, 
2, and 3.  This is easily generalized to $\nerr$ contributions to the 
systematic uncertainty and to $\nsn$ supernovae.  Such an error list, 
arising from a linear combination of independent uncertainties (similar 
to a Karhunen-Loeve transform), 
produces the wedding cake type of covariance matrix seen in 
Eq.~(\ref{eq:vsample}), and we call this the KL systematics model. 

The amplitudes of the uncertainties can be expressed either in terms of the 
individual $\sigma_j$ or in terms of the covariance matrix entries 
\beq 
\Sigma_i^2=\sum_1^i \sigma_j^2\,. \label{eq:sumsig} 
\eeq 
For definiteness, and to compare 
the results to the commonly used, diagonal LH systematic model, we 
explore the case where 
\beq 
\Sigma_i^2=\sigma_{\rm LH}^2(\langle z_i\rangle)\,. \label{eq:lhsig} 
\eeq 
For the ground-based survey we assume $\sigma_{\rm LH}(z)=0.03\,(1+z)/1.9$ 
and here $\langle z_i\rangle$ 
is the average redshift of the SNe within the range where $\sigma_i$ (but 
not yet $\sigma_{i+1}$) enters.  In the limit that $\nerr$ equals the 
number of independent error bins used in the LH model, the diagonal 
elements of the KL and LH covariance matrices are the same.

\section{Effect on Cosmology Fit \label{sec:cos}} 

We now explore the implications of the KL correlated systematic model 
on the cosmological parameter estimation.  

\subsection{Cosmology and Survey Parameters} 

We consider a survey with 2000 SNe uniformly distributed between 
$z=0.1-\zmax$, supplemented with an additional 300 SNe at $z=0.03-0.1$.  
The values of $\zmax$ considered are 0.6, 0.7, 0.8, 0.9. 
The distance modulus covariance
matrix consists of a diagonal contribution from an intrinsic dispersion 
$\sint=0.1$ for
each supernova and a correlated part given by the KL systematic model.  
We explore 
several KL scenarios with different $\nerr$ and entry redshifts 
for each error, but in all cases
the amplitude of the systematics is fixed by Eq.~(\ref{eq:lhsig}).
The statistical measurement uncertainty is taken to be subdominant.
To speed up our analysis, we combine supernovae in 0.1 redshift bins and
use the average distance modulus with variance
$\sint^2/n_i$, where  $n_i$ is the number 
of SNe in each bin. 
(Such a procedure is found to make negligible difference to the final result.) 
We include a CMB Planck prior on the distance to 
last scattering. 

The cosmological fit parameters of interest are the matter density $\om$ in 
units of the critical density, and the dark energy equation of state 
parameters $w_0$ and $w_a$ where $w(a)=w_0+w_a\,(1-a)$.  Spatial 
curvature is taken to vanish.  The fiducial model is a flat $\Lambda$CDM 
cosmology with $\om=0.28$.  The combination 
$\mathcal{M}$ of the SN absolute magnitude and Hubble constant is also 
fit.  We examine the impact of different systematic error cases on 
the dark energy 
figure of merit FOM=$1/\sqrt{\det C_{w_0w_a}}$, where $C_{w_0w_a}$ 
is the $2\times2$ submatrix of the parameter covariance matrix (inverse 
Fisher matrix). 

In Sec.~\ref{sec:errfit} we carry out a $\chi^2$ fit rather than 
a Fisher matrix analysis, finding the results agree well.  We also 
consider there the case where the $\sigma_i$ are fit simultaneously 
with the cosmology.

\subsection{Impact of Survey Characteristics \label{sec:trade}} 

A first point of interest is the behavior of the cosmological parameter 
estimation as we include various uncertainties, i.e.\ as $\nerr$ changes. 
When $\nerr=1$, then all elements of the error matrix are identical, 
corresponding to completely correlated uncertainty across all SNe 
and all redshifts; this only affects the uncertainty on the absolute 
magnitude parameter $\mathcal{M}$, not any of the cosmological parameters.  
When the off-diagonal elements are not all equal, as for $\nerr>1$, 
then the values of each $\sigma_i$ are relevant. 

\begin{figure}
  \begin{center}{
  \includegraphics[width=\columnwidth]{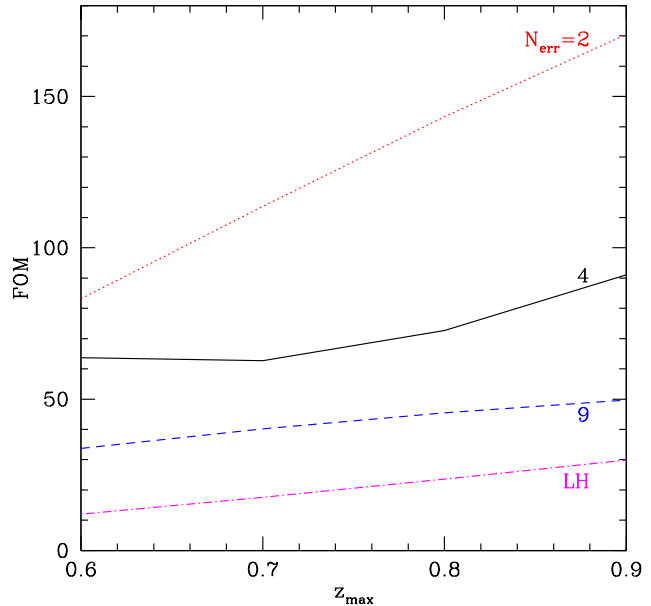}
  }
  \end{center}
  \caption{The dark energy figure of merit (FOM) is plotted vs survey 
redshift depth $\zmax$ for different cases of the global systematics 
model.  Adding more error layers, each appearing above its corresponding 
redshift, lowers the FOM for surveys extending beyond those redshifts. 
The LH local systematics model has a worse FOM because SNe in different 
redshift bins do not see correlated contributions among its 9 errors. 
}
\label{fig:zmax}
\end{figure}

The number of features contributing systematic uncertainties 
is important.  Figure~\ref{fig:zmax} shows the effect of both the number 
of error levels and the maximum depth of the survey on the FOM.

For the $\nerr=2$ case, we take the second error to come in at $z=0.1$; 
this can either reflect the highest wavelength (Ca II) spectral feature 
or be viewed as the difference between a separate, local ($z<0.1$) SN 
survey that may be carried out spectrophotometrically (e.g.\ Nearby 
Supernova Factory \cite{snf}) and a high redshift ground based SN survey 
such as Dark Energy Survey (DES \cite{des}) or LSST \cite{lsst}.  
Note that the $\nerr=9$ and LH systematics cases both have new errors 
appearing every 0.1 in redshift, and in fact have identical diagonal 
terms in the correlation matrix.  However, the global correlation causes 
more of the effect of the uncertainties to propagate into the absolute 
magnitude parameter,  
ameliorating their influence on the cosmological parameter determination, 
and so giving a higher FOM. 

We take for our canonical systematic model the case mentioned in 
Sec.~\ref{sec:corr} where the redshifting of important Si II and Ca II 
spectral features  
out of the observed wavelength range generates systematic uncertainties. 
For sensitivity diminishing at $0.9\,\mu$m these occur at $z=0.12$, 
0.47, 0.57.  This is referred to as the $\nerr=4$ case since the 
error matrix will be built up out of 4 layers, i.e.\ blocks involving 
$\Sigma_1^2$ through $\Sigma_4^2$.  The following quantitative results 
use this model. 

Most interestingly, the KL systematic, if it holds as a better description 
of SN observations for ground based observations than the uncorrelated LH 
systematic, predicts (for $\nerr=4$) a factor of 3 greater 
FOM for comprehensive surveys extending to $\zmax=0.9$.  Moreover, 
the constraints in the dark energy equation of state $w_0$-$w_a$ plane 
have slightly more complementarity with other probes such as weak 
lensing or baryon acoustic oscillations than in the LH case. 

\begin{figure}
  \begin{center}{
  \includegraphics[width=\columnwidth]{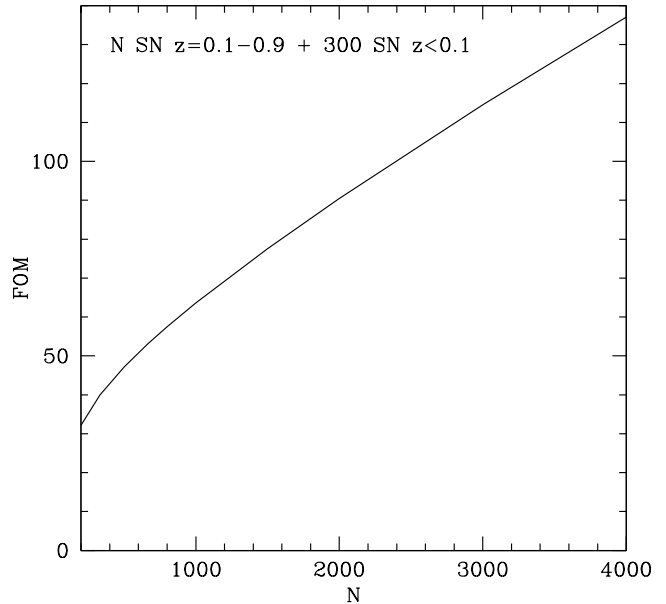}
  }
  \end{center}
  \caption{The dark energy figure of merit (FOM) is plotted vs number $N$ 
of well-characterized SNe in the high-redshift survey.  The canonical FOM 
for $N=2000$ changes by factors 1.5 or 0.7 for $N$ doubling or halving, 
respectively. 
}
\label{fig:fomn} 
\end{figure}

Considering the number $N$ of well-characterized SNe in the survey, we 
find that the FOM does not saturate with number as severely as in the 
LH systematic case.  Instead, we have 
\beq 
{\rm FOM} \sim N^{1/2}\,, 
\eeq 
over the range considered; for very large $N$ it gradually levels off. 
In the purely statistical error case FOM$\sim N$ but recall that in 
addition to these $N$ SNe at $z=0.1$--$\zmax$ we include a 
constant 300 SNe at $z=0.03$--$0.1$ and a Planck CMB prior on the distance 
to last scattering.  We emphasize that the $N$ SNe represent those with 
spectral or sufficiently high signal to noise photometric information 
to discern the different spectral feature classes down to the level of 
the $\Sigma_{\rm sys}(z)=0.03(1+z)/1.9$ systematic. 
Figure~\ref{fig:fomn} shows the FOM as a function of $N$.

\begin{figure}
  \begin{center}{
  \includegraphics[width=\columnwidth]{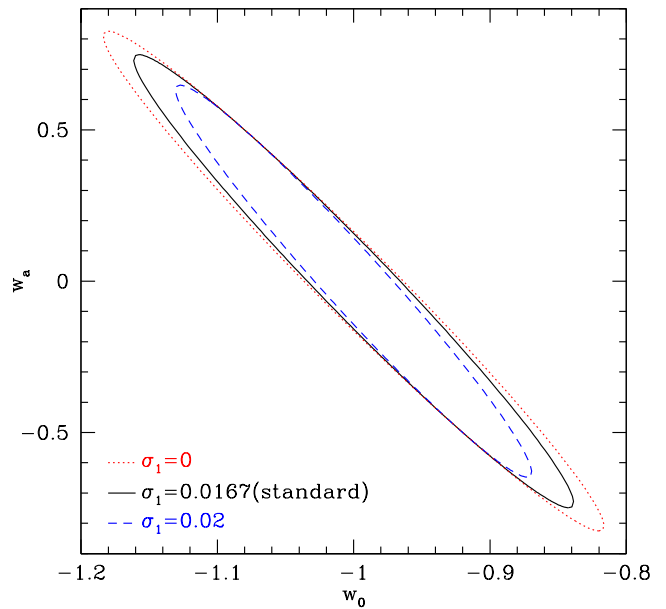}
  }
  \end{center}
  \caption{68\% confidence level contour in the dark energy equation of 
state plane is shown for varying levels of the global systematic $\sigma_1$. 
The sum $\Sigma_2^2=\sigma_1^2+\sigma_2^2$ stays fixed.  As $\sigma_1$ 
approaches 
$\Sigma_2=0.0204$, $\sigma_2\to0$ and the SNe out to $z=0.47$ share a 
common systematic, propagating mostly into $\mathcal{M}$ rather than dark 
energy parameters.  Thus the contour tightens.  Conversely, as 
$\sigma_1\to0$, the SNe become less correlated and the constraints weaken. 
}
\label{fig:sig1}
\end{figure}

Regarding the amplitude of the systematics, 
in the canonical case, the errors are 
$\{\sigma_i\}=(0.0167,0.0117, 0.0130,0.0133)$.  
The more correlated we make the SN uncertainties, by raising $\sigma_1$, 
the less the effect of the remaining errors on the 
cosmological parameters.  Figure~\ref{fig:sig1} illustrates this, where we 
adjust $\sigma_1$ to be either 0.02 (which drives $\sigma_2$ nearly to 0, 
i.e.\ $\Sigma_2\to\Sigma_1$ due to the constraint of Eq.~\ref{eq:lhsig}) 
or 0 (making $\sigma_2=0.020$).  
The area of the $w_0$-$w_a$ contour 
clearly decreases as more of the systematic becomes global, and interestingly 
the degeneracy direction rotates slightly to become more complementary to 
other probes such as weak lensing and baryon acoustic oscillations.

\begin{figure}
  \begin{center}{
  \includegraphics[width=\columnwidth]{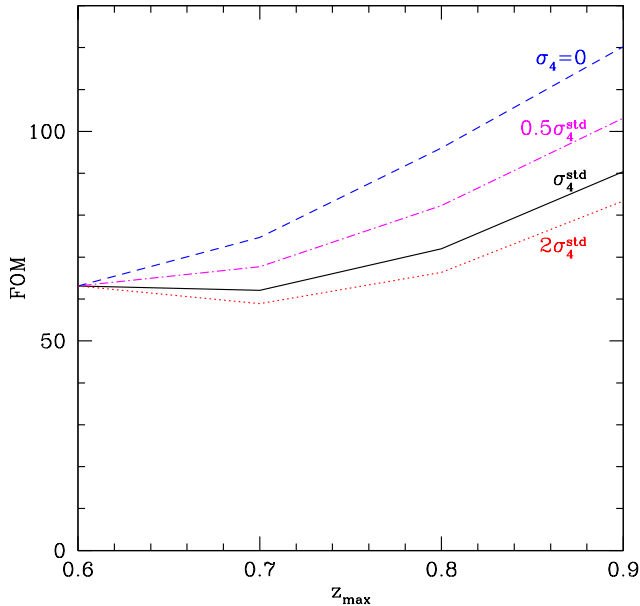}
  }
  \end{center}
  \caption{The dark energy figure of merit (FOM) is plotted for different 
values of the 
highest redshift systematic $\sigma_4$.  If the spectral feature at 
5750 \AA\ responsible for $\sigma_4$ is not actually relevant to SN 
magnitudes, so $\sigma_4\to0$, then the correlation in magnitude 
uncertainty among the highest redshift SNe grows and the FOM improves. 
Larger systematics lower the FOM at all redshifts beyond $z=0.6$ where 
this error enters.  Here $\sigma_4^{\rm std}$ is the canonical value 
0.0133 given by Eq.~(\ref{eq:lhsig}). 
} 
\label{fig:sig4}
\end{figure}

Studying the highest redshift systematic uncertainty $\sigma_4$, 
we find an analogous effect on the FOM, as seen in Fig.~\ref{fig:sig4}.  
As $\sigma_4\to0$, 
$\Sigma_4\to\Sigma_3$ and the SNe above $z=0.47$ all have a coherent 
systematic.  This improves the FOM by 33\%.  Thus, tight control of 
systematics out to the highest redshifts is beneficial.  As $\sigma_4$ is 
instead increased, the FOM worsens, but slowly, with a doubling of 
$\sigma_4$ (30\% increase in $\Sigma_4$) causing an 8\% decline in FOM. 
Since $\sigma_4$ enters at $z=0.57$, the FOMs for all values of 
$\sigma_4$ agree up to this redshift; generally the FOM then flattens 
with survey depth as the impact of the new error enters.  At even higher 
redshift the FOM gradually increases as the leverage from a larger redshift
baseline asserts itself.  Note however that the FOM can actually decrease 
with $\zmax$ if the error is large enough to overwhelm the redshift leverage.

Since the SN properties whose lack of measurement is responsible for the 
systematic uncertainties 
enter at observed wavelengths that increase for higher redshift SNe, 
an extension of the wavelength region over which the survey is sensitive 
would forestall the loss of information due to missed features.  This 
is of course one of the main reasons for considering space-based SNe 
surveys.  But within the context of ground-based surveys, if we could 
slightly extend the wavelength region over which high signal to noise 
measurements could be made, this could pay dividends for cosmological 
parameter reach. 

\begin{figure}
  \begin{center}{
  \includegraphics[width=\columnwidth]{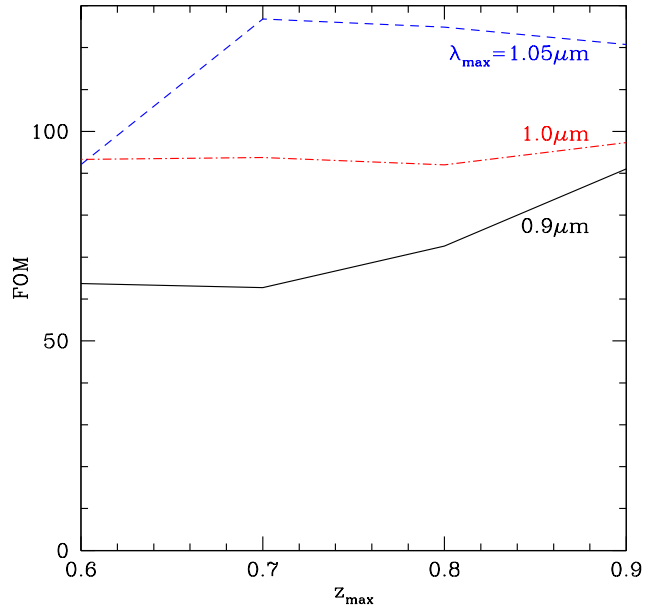}
  }
  \end{center}
  \caption{The dark energy figure of merit (FOM) is plotted vs survey
redshift depth $\zmax$ for different values of the wavelength sensitivity 
cutoff.  Increasing the wavelength range can give substantial improvements 
to FOM, even for shallower surveys. 
}
\label{fig:lamz}
\end{figure}

The ongoing development of ``extreme red'' CCDs \cite{ccd}, detectors 
with enhanced efficiency out to $1\mu$m or even $1.05\,\mu$m, offers the 
possibility of recovering from some of the observing inefficiencies caused by 
bright atmospheric emission and absorption at longer wavelengths.  
We consider these two cases in 
addition to the standard $0.9\,\mu$m cutoff.  Figure~\ref{fig:lamz} 
shows that such technology advances could have dramatic effects on 
dark energy science.

The general effect of a systematic error contribution is to suppress 
the growth of FOM with redshift at the redshift corresponding to where 
the spectral feature leaves the observational sensitivity window.  That 
is, the curves in the figure flatten at the redshifts where the errors 
enter, as mentioned above.  
For the $\lmax=0.9\,\mu$m 
case, all the errors have entered by $z=0.57$ and so the FOM curve, 
although starting low due to the accumulated systematics, increases 
with $\zmax$.  For the $\lmax=1\,\mu$m case, the errors appear at 
$z=0.25$, 0.64, and 0.74, hence the FOM curve is roughly flat over the 
range of $\zmax$ shown.  For $\lmax=1.05\,\mu$m, the errors come in at 
$z=0.31$, 0.72, 0.83 so the FOM is able to climb with redshift out to 
$\zmax=0.7$ and then levels off.  

Compared to the canonical $\zmax=0.9$ case the 
improvement in FOM by extending sensitive measurements to $\lmax=1$ or 
$1.05\,\mu$m is 7\% or 32\% respectively.  Moreover a survey with extreme 
red sensitivity to $1.05\,\mu$m is equally powerful at $\zmax=0.7$ vs 0.9 
(recall the number of SNe is kept constant at $N=2000$), giving there 
a factor of 2 improvement in FOM over a standard $0.9\,\mu$m cutoff survey.

\section{Further Applications \label{sec:other}}

\subsection{Error Fitting \label{sec:errfit}} 

Systematics represent the residuals after known errors are taken into 
account.  One usually estimates, either empirically through training 
and validation subsets or using simulation and modeling guidance, the 
level of the systematic.  An example is the 0.03 mag amplitude in the LH 
model.  In Sec.~\ref{sec:cos} we assume that those levels are
already well-characterized when incorporated in the distance modulus
covariance matrix.  Alternatively, within the cosmology
fit one can simultaneously estimate the level of variance or
introduce new parameters to fit out the effect of the systematic. 

To explore 
this we carry out simultaneous fits to various systematics model 
parameters along with the cosmological parameters.  
In place of the Fisher analysis, we use MINUIT \cite{minuit} 
for parameter fitting and the 
resultant Hessian matrix to calculate the FOM.  Contributions to 
the fit likelihood come from not only the usual $\chi^2$ term but also 
a penalty term $\ln\det C$ involving the covariance matrix $C$.  For 
example, this approach was carried out to fit the intrinsic SN dispersion 
$\sint$ by \cite{alexpasp}.  We do not bin but analyze the full 
uniformly-distributed 2000 SNe
plus the 300 local SNe.  Since we are interested in fit
uncertainties, when possible we do not
realize dispersion on the simulated data to ensure that the
best fit and input parameters are identical.
We restrict these studies to the $\nerr=4$, $\zmax=0.9$ case.

We first check that this likelihood approach gives the same results 
as the Fisher analysis for our canonical case.  The FOMs agree 
to better than 5\%. 

Next, we consider that the estimate of the absolute magnitude may have
a different intrinsic dispersion depending on the number of spectral
features used in its estimate.  This is done by considering
the $\sigma^2_{{\rm int},n_a}$ in Eq.~(\ref{MCov:eqn}) as (four) fit 
parameters rather 
than assuming they equal 0.1.  Since we are fitting for variances,
we must instantiate dispersion in the simulated data.  We perform only
one instantiation as a single fit is time consuming
with a  $2300 \times 2300$ matrix inversion in each likelihood evaluation. 
The uncertainty found for each $\sigma_{{\rm int},n_a}$ 
is $\sim {\rm few}\times 10^{-2}$, the level to which we can detect 
real differences in the absolute magnitude models.  The FOM is 
within 3\% of the fixed $\sint=0.1$ case. 

The correlated systematics $\sigma_i$ reflect the residual
uncertainties in the absolute magnitude offsets in redshift bins
defined by spectral-feature visibility.  These $\nerr$ offsets can be
fit for simultaneously with the cosmology since the luminosity
distance model is smooth, without sudden steps (see the next
subsection for the case of a smooth magnitude drift).  Fitting for the
biases on $M_i$ (without the $V$ term in the covariance matrix), the
realizations of $\delta M_i$ are determined to $\sim 10^{-2}$ and the
FOM falls to 59 (i.e.\ by $\sim40\%$).  However, when adopting the
$\sigma_i$ given by Eq.~(\ref{eq:lhsig}) as priors on $\delta M_i$
(with mean 0), the realizations of $M_i$ are determined to $\sim
7.5\times10^{-3}$ and the FOM improves to 102 (by 12\%).  The smooth Hubble
diagram of the cosmological parameters allows the self-calibration of
magnitude offsets, leading to an improved FOM over the canonical case.

\subsection{Population Drift \label{sec:pop}} 

As used so far, the KL model assumes that once information on a 
spectral feature is lost, the uncertainty enters and remains constant with 
redshift.  In terms of a magnitude correction based on that feature, 
this makes sense: no information means no correction and hence the 
added uncertainty.  But we can also consider the spectral feature 
as aiding classification of subtypes of SNe Ia with different absolute 
magnitudes.  

The various subtypes will all combine to give some average 
absolute magnitude for the sample.  With the loss of information, 
the proportion of subtypes is unlikely to suddenly change abruptly, 
and so the average should initially remain a good approximation.  At 
higher redshifts, however, the population fractions may drift further 
and so the previous average becomes increasingly uncertain as an 
estimate of the true sample average.  In this population drift 
interpretation, the systematic uncertainty increases as the redshift 
increases beyond that where the feature is lost. 

We explore a modified KL model where the systematic scales as 
\beq 
\sigma_i(z)=\sigma_{i\star}\,\frac{z-z_i}{z_\star-z_i} 
\eeq 
in the error table $S$.  The exception is $\sigma_1$ which we 
take to be a global uncertainty for SN standardization.  
The covariance matrix then contains elements 
\begin{equation}
\begin{split}
\left<\delta M_a\,\delta M_b\right>&= \sigma_1^2 \\ 
& \quad +\sum_{i=2}^{N-\min{(n_a,n_b)}+1} 
\sigma_{i\star}^2
\left(\frac{z_a-z_i}{z_\star-z_i}\right)
\left(\frac{z_b-z_i}{z_\star-z_i}\right), 
\end{split}
\end{equation} 
where $z_a$, $z_b$ are the redshifts of the two SNe being correlated 
and $z_i$ is the entry redshift of the systematic $\sigma_i$. 

The systematic $\sigma_i$ is thus not a constant entry in a wedding 
cake layer but rather gives a ramp up: it starts at 0 when it enters 
at $z_i$ and attains an amplitude 
$\sigma_{i\star}$ at $z=z_\star=0.9$.  For comparison 
we choose $\sigma_{i\star}$ to equal the $\sigma_i$ as given 
for the redshift independent, or flat, KL model in the earlier part of 
the paper to match the LH model amplitudes. 
Due to the scaling with $z$, the systematic amplitude will be less 
than the flat KL case for all $z<z_\star$.  Not surprisingly, therefore 
the FOM for the scaling case is larger (by 28\% in the canonical 
$\nerr=4$, $\zmax=0.9$ case).  This implies that 
such population drift systematics have less effect (for the same 
amplitude at the maximum redshift) than the magnitude correction 
systematic.

\section{Conclusions \label{sec:concl}} 

Systematics are the ultimate limiting factor in any cosmological 
probe.  We show that uncertainties due to the loss of information of 
supernova spectral features can correlate distance errors for all 
redshift higher than where the information is lost.  One expects such 
loss in ground-based surveys where sensitivity declines strongly beyond 
9000 \AA.  

Such correlation in the error model 
can actually have a tempering effect, relative to a localized 
uncertainty such as in the LH model, in that a wider range of supernovae 
experience the effect, it propagates mostly into the absolute 
magnitude determination, and so the impact on the cosmological 
parameter estimate is reduced.  For the same level of dispersion 
(autocovariance), the KL correlated systematic model allows a larger 
dark energy figure of merit and slower approach to the systematic floor. 

Extending the wavelength sensitivity into the near infrared, such as 
may be possible with new technology ``extreme red'' CCDs that also 
cover the optical region, has science potential that should be considered 
in mission designs.  For a 
survey depth of $\zmax=0.9$ the gain in figure of merit is 32\% on 
going out to $1.05\,\mu$m.  Moreover, the same FOM can be achieved 
even if the survey depth is $\zmax=0.7$, since the added systematics 
beyond $z=0.72$ due to loss of Si II features counteract the increased 
redshift baseline, for a fixed number of supernovae. 

Viewing spectral information loss as opening the analysis 
to uncorrected population drift, we find that this does not have 
as much impact as for the magnitude correction ansatz.  
As redshift increases, information might also be gained from rest 
frame UV features, but this would occur at low redshift ($z\approx0.1$-0.25) 
and is not expected to have much impact on survey depth design.  Currently 
the situation concerning UV features for magnitude correction, or their 
ability to substitute for the higher wavelength information discussed 
here, is not clear.  

The calculations here are meant illustratively, to show the method and 
possible impacts of correlated systematics.  We do not claim that the 
Branch et al.\ model \cite{branch06} for supernova subclassification is 
truth, that those particular wavelengths carry the essential information 
on supernova magnitudes.  For example, Bailey et al.\ \cite{bailey} have 
identified certain distinct series of spectral flux ratios and ranked 
their correlation with absolute magnitude. 
Nevertheless, the approach of analyzing correlated systematics, and the 
benefit of extending the wavelength range, 
e.g.\ through red sensitive CCDs, should be general.

\acknowledgments

We acknowledge useful discussions with David Rubin.  AK thanks the 
Institute for the Early Universe for hospitality.  
This work has been supported in part by the Director, Office of Science, 
Office of High Energy Physics, of the U.S.\ Department of Energy under 
Contract No.\ DE-AC02-05CH11231, and the World Class University grant 
R32-2009-000-10130-0 through the National Research Foundation, Ministry 
of Education, Science and Technology of Korea.


\begin{thebibliography}{99}

\bibitem{systematicspapers} 
P. Astier, J. Guy, R. Pain, \& C. Balland, \aap, 525, A7 (2011).\\ 
L. Faccioli et al., Astropart. Phys. accepted [arXiv:1004.3511]\\ 
A.~G. Kim, E.~V. Linder, R. Miquel, \& N. Mostek, \mnras, 347, 
909 (2004).\\ 
M.~C. March, R. Trotta, L. Amendola, \& D. Huterer, 
arXiv:1101.1521 (2011).\\ 
J. Nordin, A. Goobar, J. J{\"o}nsson, \jcap, 2, 8 (2008).\\ 
N. Mostek, Ph.D.~Thesis (2007). 
http://sites.google.com/a/lbl.gov/njmostek/docs/PhD.pdf\\ 
J. Samsing, \& E.~V. Linder, \prd, 81, 043533 (2010). 

\bibitem{coping} 
D. Branch, S. Perlmutter, E. Baron, P. Nugent, 
arXiv:astro-ph/0109070

\bibitem{linlike} 
E.~V. Linder, Phys. Rev. D 79, 023509 (2009).
[arXiv:0812.0370] 

\bibitem{kimmiq} 
A.~G. Kim \& R. Miquel, Astropart. Phys. 24, 451 (2006). 
[arXiv:astro-ph/0508252] 


\bibitem{facc}
L. Faccioli et al., Astropart. Phys. accepted 
[arXiv:1004.3511]

\bibitem{linhut} 
E.~V. Linder \& D. Huterer, Phys. Rev. D 67, 081303 (2003). 
[arXiv:astro-ph/0208138]

\bibitem{vernier} 
E.~V. Linder, Phys. Rev. D 75, 063502 (2007). 
[arXiv:astro-ph/0612102] 

\bibitem{branch06} D. Branch et al.\ 
PASP 118, 560 (2006). [arXiv:astro-ph/0601048] 

\bibitem{spectraldiversity} 
V. Arsenijevic, S. Fabbro, A.~M. Mour{\~a}o \& A.~J. Rica da Silva,  \aap, 492, 535 (2008).\\ 
S. Bailey et al., \aap, 500, L17 (2009).\\ 
S. Blondin, K.~S. Mandel, \& R.~P. Kirshner, \aap, 526, A81 (2011).\\ 
S. Benetti, et al., \apj, 623, 1011 (2005).\\ 
R.~S. Ellis et al., \apj, 674, 51 (2008)\\ 
R.~J. Foley et al., \apj, 684, 68 (2008).\\ 
J. Nordin, et al, \aap, 526, A119 (2011).\\ 
X. Wang, et al. \apjl, 699, L139 (2009).

\bibitem{snf} 
Nearby Supernova Factory http://snfactory.lbl.gov 

\bibitem{des}
Dark Energy Survey http://www.darkenergysurvey.org/

\bibitem{lsst} 
Large Synoptic Survey Telescope http://www.lsst.org/lsst

\bibitem{ccd} 
C. Bebek, private communication 

\bibitem{minuit} 
Minuit http://www.cern.ch/minuit 

\bibitem{alexpasp} 
A. Kim, PASP, in press  
[arXiv:1101.3513] 

\bibitem{bailey} 
S. Bailey et al., Astron. Astrophys. 500, L17 (2009).
[arXiv:0905.0340] 


\end{thebibliography}
\end{document}